\documentclass[%
 reprint,
superscriptaddress,
 amsmath,amssymb,
 aps,
prb,
]{revtex4-2}
\usepackage{libertine}
\usepackage{graphicx}
\usepackage{dcolumn}
\usepackage{bm}
\usepackage{xcolor}

\usepackage{comment}

\begin{document}

\preprint{APS/123-QED}

\title{Optical excitation of bulk plasmons in n-doped InAsSb thin films :  investigating the second viscosity in electron gas}

\author{Antoine Moreau}%
\email{antoine.moreau@uca.fr}
\affiliation{Universit\'e Clermont Auvergne, CNRS, Institut Pascal, F-63000 Clermont-Ferrand, France}%
\author{\'Emilie Sakat}
\affiliation{Universit\'e Paris Saclay, Center for Nanoscience and Nanotechnology, C2N UMR9001, CNRS, 91120 Palaiseau, France}
\author{Jean-Paul Hugonin}
\affiliation{Universit\'e Paris-Saclay, Institut d'Optique Graduate School, CNRS, Laboratoire Charles Fabry, 91127 Palaiseau, France
}%
\author{Téo Mottin}
\affiliation{Universit\'e Clermont Auvergne, CNRS, Institut Pascal, F-63000 Clermont-Ferrand, France}%
\author{Aidan Costard}
\affiliation{Universit\'e Clermont Auvergne, CNRS, Institut Pascal, F-63000 Clermont-Ferrand, France}%
\author{Sarah Abdul-Salam}
\affiliation{Universit\'e Clermont Auvergne, CNRS, Institut Pascal, F-63000 Clermont-Ferrand, France}%
\author{Denis Langevin}
\affiliation{Universit\'e Clermont Auvergne, CNRS, Institut Pascal, F-63000 Clermont-Ferrand, France}%
\author{Patricia Loren}
\affiliation{Institut de Physique de Nice, CNRS, Université Côte d'Azur, 06200 Nice, France}%
\author{Laurent Cerutti}
\affiliation{Institut d'Electronique et des Systèmes, CNRS, Université de Montpellier, 34095 Montpellier, France}%
\author{Fernando Gonzalez Posada Flores}
\affiliation{Institut d'Electronique et des Systèmes, CNRS, Université de Montpellier, 34095 Montpellier, France}%
\author{Thierry Taliercio}
\email{thierry.taliercio@umontpellier.fr}
\affiliation{Institut d'Electronique et des Systèmes, CNRS, Université de Montpellier, 34095 Montpellier, France}%
\date{\today}

\begin{abstract}
We demonstrate that including the second viscosity of an electron gas in the hydrodynamic model allows for highly accurate modeling of the optical response of heavily doped semiconductors. In our setup, which improves resonance visibility compared to previous approaches, plasmon resonances become more distinct, allowing for a detailed analysis of the underlying physics. With advanced fitting techniques based on a physics-informed cost function and a tailored optimization algorithm, we obtain a close agreement between simulations and experimental data across different sample thicknesses. This enhanced resonance visibility, combined with our integrated approach, shows that key parameters such as doping level and effective electron mass, as well as the second viscosity of the electron gas, can be retrieved from a single optical measurement. The spatial dispersion taken into account in the hydrodynamic framework is essential for accurately describing the optical response of plasmonic materials in this frequency range and is likely to become a standard modeling approach.
\end{abstract}

\maketitle

Plasma oscillations in conducting materials were first described by Tonks and Langmuir\cite{tonks1929oscillations}, who identified these longitudinal waves, similar to sound, now called plasmons, in ionized gases near the plasma frequency. Ferrell\cite{ferrell1962plasma} later predicted that thin metal films should show resonances just above the screened plasma frequency, where the permittivity approaches zero. Melnyk and Harrison\cite{melnyk1968resonant,melnyk1970theory} used a hydrodynamic model that included electron-electron interactions to describe the metallic response more accurately. They predicted that metal slabs should support several odd-order bulk plasmon resonances, not just the single resonance that Ferrell had predicted. These predictions were confirmed in very thin potassium films\cite{anderegg1971optically}, but for most metals the plasma frequency sits right in the middle of interband transitions, where absorption is so strong that plasmonic resonances cannot exist.

In heavily doped semiconductors, however, the plasma frequency is in the infrared range, far from interband transitions. The community first used this to develop a practical tool: the main plasmon resonance (often called the Ferrell or Epsilon-Near-Zero resonance) can be used with calibration charts to measure doping concentrations from simple optical measurements\cite{taliercio2014brewster}. This application was developed without much concern for the theoretical predictions about higher-order resonances.

Separately, experiments on particle-on-mirror systems showed that Drude's model\cite{drude1900elektronentheorie} is not sufficient to accurately describe particles on mirror resonances\cite{Ciraci2012,reynaud2018enhancing}, but that the hydrodynamic model was able to do so. This has renewed interest in the hydrodynamic model, for which improvements have been proposed\cite{Moreau2013,mortensen2014generalized}. It has also been implemented in various numerical simulation methods\cite{Toscano2012,benedicto2015numerical,schmitt2016dgtd,kwiecien2023nonlocal,langevin2024pymoosh}, making it more widely usable.

Recently, De Ceglia et al.\cite{de2018viscoelastic} showed that very thin films of heavily n-doped semiconductors do exhibit multiple bulk plasmon resonances, confirming what the hydrodynamic model predicted decades ago. This has renewed interest in properly understanding these effects. Quantum mechanical calculations can describe these systems accurately\cite{vasanelli2020semiconductor}; however, it seems that the hydrodynamic model offers a practical alternative that captures the main physics. 

In this work, we show that bulk plasmon resonances can be observed in doped semiconductor films that are an order of magnitude thicker than those studied previously. We use a pseudo-ATR configuration where the semiconductor sits between two high-index materials, which makes the resonances much easier to observe. We find that including the bulk viscosity (or second viscosity) of the electron gas in the hydrodynamic model is necessary for accurate fits. This provides a natural explanation for why the nonlocal parameter needs an imaginary part\cite{halevi1995hydrodynamic,mortensen2014generalized,de2018viscoelastic}. Using a global optimization algorithm combined with a physics-informed cost function, we  extract material parameters —doping concentration, effective mass, and bulk viscosity— from a single optical measurement without the need for calibration charts. Our fitting procedures and analysis tools are available open-source, making this characterization technique immediately accessible to the community.

In the first part of this paper, we review the basics of spatial dispersion in an electron gas and the characteristics of plasmons to provide a foundation for a thorough understanding of our results. We then detail our experimental findings, demonstrating how the observed resonance positions align with predictions from the hydrodynamic model. In the third section, we explain how the use of a global optimization algorithm and an adapted cost function to fit our experimental data allows us to retrieve all the model parameters.

\section{Theoretical framework}

\subsection{Drude model context}

Drude's model is first based on the idea that the volume current density $\mathbf{j}$ can be integrated as a screened polarization $\mathbf{P_f}$ of the medium, using the relation $\mathbf{j}=\frac{\partial \mathbf{P_f}}{\partial t}$. A metal or a doped semi-conductor with a free carrier gas can thus always be described with such an effective polarization. In the framework of the Drude model, the link between this polarization and the electric field can be written
\begin{equation}\label{eq:drude}
   \frac{\partial^2 \mathbf{P_f}}{\partial t^2}+\gamma \frac{\partial \mathbf{P_f}}{\partial t}=\varepsilon_0 \omega_p^2 \mathbf{E},
\end{equation}
where $\gamma$ accounts for the friction of the electron gas against the lattice and $\omega_p^2 = \frac{n e^2}{m^* \varepsilon_0}$, $n$ being the density of the electron gas and $m^*$ the effective mass of electrons.

It should be underlined that the effective polarization, in the harmonic regime (we assume a $e^{-i\omega t}$ time dependency here), is actually proportional to the displacement of the electrons (or carriers more generally) with respect to their average position. The second term is linked to the friction of the electrons on the lattice, while the last is the electric force. In the harmonic regime, the displacement of the electrons is always opposite to the electric force because there is no other restoring force. The effective polarization is, thus, opposite to the electric field and dominates if the frequency is low enough, making the permittivity negative. 

The rest of the medium is characterized by a susceptibility $\chi_b$ that may depend on the frequency. In the framework of the Drude model, the permittivity of the material is thus  given by 
\begin{equation}
    \varepsilon = \varepsilon_0 \left(1 + \chi_b - \frac{\omega_p^2}{\omega^2 + i\gamma \omega}\right).\label{eq:permittivity}
\end{equation}

The screened plasma frequency, for which the real part of the permittivity vanishes and which is the only frequency that can be measured directly,  is given approximately by 
\begin{equation}
    \omega_{0} = \frac{\omega_p}{\sqrt{1 + \chi_b}}.
\end{equation}

It is referred to as the screened plasma frequency or the epsilon-near-zero (ENZ) frequency because, below this frequency, the semiconductor can be considered a plasmonic material, similar to metals in the visible range. Although many authors use "plasma frequency" to describe the ENZ frequency, these two frequencies must be distinguished, particularly given the high susceptibility of semiconductors in the infrared range (typically, $\chi_b \simeq 10$).

\vspace*{0.5cm}

\subsection{Including second viscosity}

Drude's model is not sufficient to describe and understand plasmons, as it is necessary to take into account electron-electron repulsion in the description of the electron gas. The simplest way to do so is to consider the electron gas as a Newtonian fluid and to incorporate a supplementary pressure term into the fluid equation\cite{Ciraci2013}. After linearization, these equations are coupled with Maxwell's equations to yield a coherent description of the optical response of the electron gas. We will follow this path here, but we will consider all viscosity terms \cite{ash1991second,gad1995stokes,graves1999bulk,buresti2015note}. The equation governing the electron gas dynamics in this framework is 
\begin{widetext}
\begin{equation}
n\, m^*\, \frac{\partial \mathbf{v}}{\partial t} + n\, m^* \, \mathbf{v} \cdot \nabla \mathbf{v} = - \nabla p + \mu \Delta \mathbf{v} + (-e)\, n \,\mathbf{E} -n\,m^* \,\gamma \,\mathbf{v} +  \xi\, \nabla (\nabla \cdot \mathbf{v})
\end{equation}
\end{widetext}
where $n$ and $m^*$ are the electron density and electron effective mass, respectively; $\mathbf{v}$ is the electron velocity, $p$ is the pressure inside the electron gas, $\mu$ is the (shear) viscosity, $\gamma$ is a damping factor, and $\xi$ is the second (or bulk) viscosity\cite{graves1999bulk}. 

In the framework of the hydrodynamic model, viscous terms are usually not included\cite{Moreau2013,Ciraci2013,mortensen2014generalized}. It recently became clear, however, that electron gases do exhibit a viscous behavior\cite{bandurin2016negative,polini2020viscous}. The shear viscosity has been integrated into the hydrodynamic model in a recent work for the first time\cite{de2018viscoelastic}. The second viscosity term is usually assumed to be negligible because this form of viscosity becomes relevant only when the fluid expands or contracts significantly. It is typically neglected when deriving the Navier-Stokes equations because most fluid flows are essentially incompressible, even if the fluid itself is very compressible, leaving shear as the only source of friction\cite{buresti2015note}. This assumption, called the Stokes hypothesis, is so common and accurate that the bulk viscosity of a gas is challenging to estimate. Furthermore, monoatomic gases are expected to have a vanishing bulk viscosity. Electrons, also lacking internal degrees of freedom, might be expected to behave similarly. It would thus seem natural to neglect the second viscosity for electron gases.

However, when plasmons are excited, as they involve periodic compression and rarefaction of the electron density, with minimal shear deformation, bulk viscosity naturally dominates over shear viscosity. This is inspired by what occurs in classical gases, where bulk viscosity is the primary mechanism for sound attenuation and typically exceeds shear viscosity by orders of magnitude \cite{ash1991second,cramer2012numerical}. We therefore adopt this perspective for what follows, keeping only the bulk viscosity contribution to the hydrodynamic model.


We now linearize the equations to obtain a linear response model. The velocity field $\mathbf{v}$ is considered a first order term, so that the inertial term $\mathbf{v} \cdot \nabla \mathbf{v}$ can be neglected also. The electron gas density $n$ can be decomposed into a constant term $n_0$ and a first-order fluctuation $n_1$ so that finally we can write $- n_0 e \partial_t \mathbf{v} \simeq \partial_t\, \mathbf{j}$ at the first order. This can also be  done for the spatial derivatives of the bulk viscosity term. Taking these steps into account leads to
\begin{equation}
     -\frac{m^*}{e}\, \partial_t \mathbf{j} = -\nabla p + \frac{m^*}{e} \,\gamma\, \mathbf{j} - n_0 \,e \,\mathbf{E} -\frac{\xi}{n_0\,e}\, \nabla (\nabla \cdot \mathbf{j})
\end{equation}

The pressure term is given by quantum theories and is a power of $n$ (usually $n^{3}$ is retained \cite{halevi1995hydrodynamic}). Whatever the exponent, the gradient of the pressure is simply proportional to the gradient of the electron gas density $n$. The nonlocal parameter $\beta$ is introduced at this stage so that 
\begin{equation}
    -\nabla p =- m^* \beta^2\, \nabla n.
\end{equation}

The parameter $\beta$ quantifies the non-local effects and is called the hydrodynamic parameter. It is usually defined as $\beta=\sqrt{3/5}\,v_F$ with $v_F=\frac{\hbar}{m^*}(3 \pi^2)^{1/3}n_0^{1/3}$, the Fermi velocity \cite{Ciraci2013}. The coefficient between $\beta$ and $v_F$ actually depends on the frequency. However, close to the screened plasma frequency $\sqrt{3/5}$ is a very accurate assumption \cite{halevi1995hydrodynamic}.

The continuity equation can be written
\begin{equation}
    \nabla . \mathbf{j} = - \partial_t (-e\,n) = e\,\partial_t n_1.
\end{equation}

Since  $\mathbf{j} = \partial_t \mathbf{P_f}$, this equation can be integrated to yield $\nabla \cdot \mathbf{P_f} = e n_1 $. The pressure term can be finally written
 \begin{equation}
     -\nabla p = -\frac{m^*}{e} \beta^2\, \nabla(\nabla \cdot\mathbf{P_f}).
 \end{equation}
This term obviously includes spatial derivatives of the polarization, making the description non-local. We underline that the continuity equation can be written $\nabla.\mathbf{P_f} = -\rho$ in this framework, so that this supplementary term truly appears as a force pushing electrons away from any concentration of negative charges.
For the second viscosity term, we have:
\begin{equation}
-\frac{\xi}{n_0 e} \, \nabla (\nabla \cdot \mathbf{j}) = -\frac{\xi}{n_0 e} \, \partial_t \nabla (\nabla \cdot \mathbf{P_f}),
\end{equation}
which is simply proportional to the time derivative of the pressure term. In the harmonic regime, this gives:
\begin{equation}
-\left(\beta^2 - \frac{i \omega \xi}{n_0 m^*} \right) \nabla (\nabla \cdot \mathbf{P_f}) - (\omega^2 - i \omega \gamma )\mathbf{P_f} = \epsilon_0 \omega_p^2 \mathbf{E}.
\end{equation}
Thus, the second viscosity term introduces a frequency-dependent imaginary contribution to $\beta^2$. Following the notation of previous works\cite{mortensen2014generalized} we can define a complex parameter 
\begin{equation}
    \eta^2 = \beta^2 - i\,\omega \frac{\xi}{n_0 m^*}\label{eq:eta}
\end{equation}
 which naturally incorporates bulk viscosity effects.

The inclusion of bulk viscosity thus provides a physically grounded explanation for the imaginary part of $\beta^2$ that has been empirically necessary to fit experimental data\cite{Raza2013}. Unlike the GNOR\cite{mortensen2014generalized} model's additional diffusion currents -- which raise questions about charge conservation -- or shear viscosity -- which produces spurious resonances\cite{raza2011unusual} -- bulk viscosity naturally yields an imaginary part proportional to $\omega$ through the physical compression of the electron gas. This suggests that the successful GNOR phenomenology in fact captures the effects of bulk viscosity, making the distinction between the GNOR and hydrodynamic models unnecessary.

\subsection{Plasmons}

Having established the hydrodynamic model with bulk viscosity, we now examine the nature of the waves that can propagate in such a nonlocal medium.

In a medium containing an electron gas, the electric field can be decomposed as the sum of a divergence-free component (the transverse wave) and a curl-free component (the longitudinal wave)\cite{bhatia2012helmholtz}. Since the divergence of the magnetic field is always null, the magnetic field always belongs to the purely transverse part of the decomposition. 

The two kinds of waves evolve independently from each other -- in the sense that the transverse wave is never converted into a longitudinal one along its propagation in a homogeneous medium. The transverse wave is completely insensitive to spatial dispersion, it is described as a wave propagating in a local medium with a permittivity given by the Drude model so that we call it light in the following. We will continue to call the longitudinal wave plasmon, though technically, a plasmon is the quantum corresponding to this wave. 

We underline generally that while classical electromagnetic theorems can be extended to spatially dispersive media, their generalization is not straightforward\cite{sakat2021generalized}.

At an interface between the non-local medium and any local medium, because the electron gas can not escape from the material in which it is contained, the component of the current perpendicular to the interface vanishes. This is simply given in the harmonic regime by $\mathbf{P_f}\cdot\mathbf{n}=0$ and provides the necessary additional boundary condition\cite{Moreau2013,burda2023nonlocal} required by the hydrodynamic model to be complete. 

When light from a local medium illuminates an interface, the two types of waves (transverse and longitudinal) are excited within the nonlocal medium. The boundary conditions determine the amplitude of each of these waves. More specifically, the strength of the longitudinal wave excitation depends on the direction of the electric field at the interface. At normal incidence, with the electric field purely tangential to the surface, the plasmon remains unexcited, but as the incidence angle increases, the excitation of the longitudinal wave becomes stronger.

When a longitudinal wave inside a nonlocal material encounters an interface with a local material,
it is reflected, but a transverse wave is also excited in the process. This phenomenon is best described using a scattering matrix formalism\cite{benedicto2015numerical} adapted to the hydrodynamic model, which we employ here to calculate the reflectance of any given structure, assuming a complex $\beta^2$ parameter.

The dispersion relation of the plasmon\cite{Moreau2013}, taking into account losses due to friction with the lattice and the second viscosity, can be written as:
\begin{equation}
    \omega^2 +i\left(\gamma + \frac{\xi \,\mathbf{k}^2}{n_0\,m^*}\right) \omega = \omega_0 + \beta^2 \mathbf{k}^2.
\end{equation}

while the dispersion of light inside the nonlocal medium can be written as:
\begin{equation}
    \omega^2  =  \omega_0^2 + \frac{c^2}{1+\chi_b} \mathbf{k}^2.
\end{equation}

As the Fermi velocity and thus $\beta$ is much smaller than the speed of light, this means that, above $\omega_0$, the wavevector of plasmons is a few orders of magnitude larger than that of light (whether in vacuum or inside the medium).

\begin{figure}
    \centering
    \includegraphics[width=\linewidth]{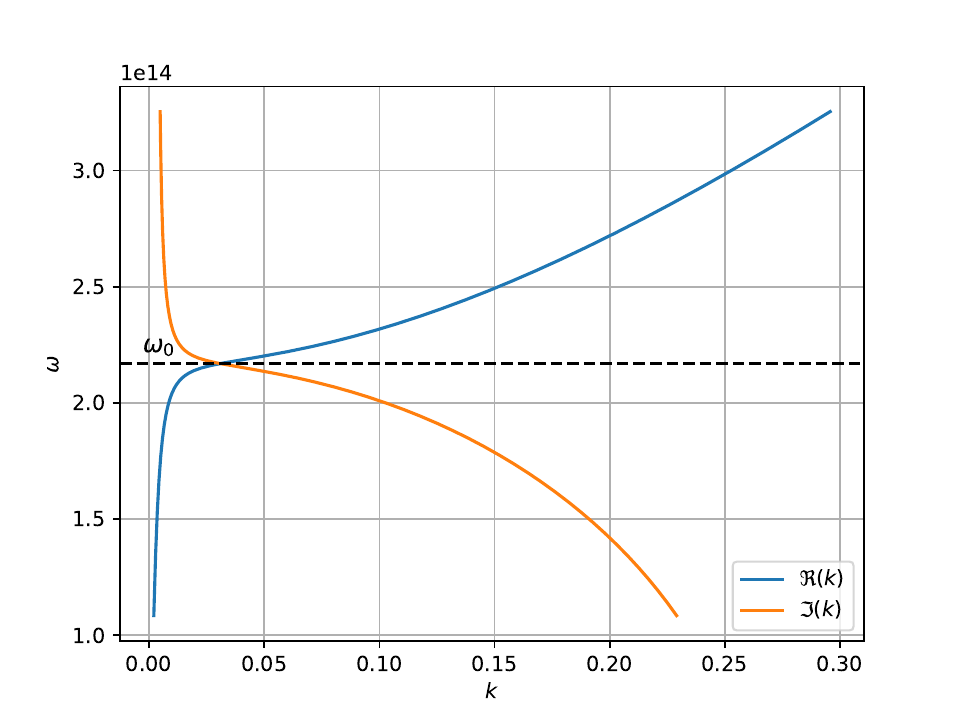}
    \caption{Dispersion curve for the plasmon with $\omega_0$=2.169.10$^{14}$ rad.s$^{-1}$, $\gamma$=5.901.10$^{12}$ rad.s$^{-1}$, $\beta = 8.2019\times 10^{5}$ m.s$^{-1}$. For $\omega >\omega_0$ the plasmon becomes propagative, showing a wavevector that is essentially real, while it is evanescent for $\omega<\omega_0$ with a dominant imaginary part.}
    \label{fig:dispersion}
\end{figure}

An example of dispersion curve for the plasmon is shown in Figure~\ref{fig:dispersion}. The parameters chosen here are in agreement with the experiments and data fits presented in the following. Below the screened plasma frequency $\omega_0$, plasmons are evanescent, just as light. Their wavevector being essentially imaginary, it is possible to define a typical penetration depth $\delta_{\mathrm{p}}$ for these waves. It is usually much shorter for plasmons than for light. All accounted for, this typical length is proportional to $n_0^{-1/6}m^{*-1/2}$, thus decreasing with the electronic density. In highly-doped semiconductors the free carrier density $n_0$ is orders of magnitude lower than in noble metals and the effective mass $m^{*}$ is also lower -- this explains why $\omega_\mathrm{p} = \sqrt{\frac{n_0 e^2}{m^* \varepsilon_0}}$ is in the infra-red range.

However, these considerations show that theoretically, the plasmon skin depth $\delta$ is at least 5 times larger or more in highly-doped semiconductors than in noble metals. This constitutes another indication that non-local effects can be expected to play a much larger role in the semiconductors' response, even below the screened plasma frequency.

\subsection{Resonances in a metallic slab}

Within the framework of the Drude model, when a horizontal slab containing a free electron gas is illuminated in normal incidence, the gas is free to oscillate horizontally. No resonance is associated with such an oscillation. When the angle of incidence is different from zero, a vertical oscillation is predicted to occur at exactly the screened frequency $\omega_0$ when the electric field penetrates the entire slab. For a rigid gas to oscillate vertically, charges must accumulate on both surfaces of the slab, creating a restoring force that drives the oscillation and induces a resonance -- a mechanism suggested by Ferrell\cite{ferrell1962plasma}. Shortly thereafter, a resonance was observed more or less in the predicted conditions\cite{steinmann1960experimental} and it is sometimes said that it corresponds to the excitation of a so called Brewster mode. There is however no guided or cavity mode that is excited in that case. The nature of the resonance is fundamentally different from a cavity resonance for light, or of the excitation of a guided mode.

When interactions between electrons are considered, surface charges can no longer be described as such; instead, they appear as volume charges near the surface, a phenomenon known as 'smearing' of the charges. The electron gas can no longer be viewed as a rigid entity moving as a whole; rather, it supports plasmons. When a plasmon is excited by incident light, the wavevector component along the interface is conserved. However, the wavevector of the plasmon being much larger than that of light, it is dominated by its vertical component, regardless of the incidence angle. The slab can therefore be treated as a cavity for these waves, with resonances occurring at relatively small thicknesses as shown by Melnyk and Harrison\cite{melnyk1968resonant,melnyk1970theory}.

With that picture in mind, it is easy to estimate the position of the resonances. Whatever the incidence angle, the plasmon can be considered to propagate almost perfectly perpendicularly to the interfaces so that the resonance condition for such a cavity can be expressed as:
\begin{equation}
    k = \frac{ \ell \pi}{h}
\end{equation}
where $\ell$ is a non-zero integer, $h$ the thickness of the slab and $k$ the plasmon wavevector. 

Using the lossless dispersion relation for the plasmon, this provides an approximation for the resonance frequencies:
\begin{equation}
    w_m = \sqrt{w_0^2 + \left(\frac{\ell \beta \pi}{h}\right)^2}.\label{eq:cavity}
\end{equation}

\begin{figure}[h!]
\centering
{\includegraphics[width=0.9\linewidth]{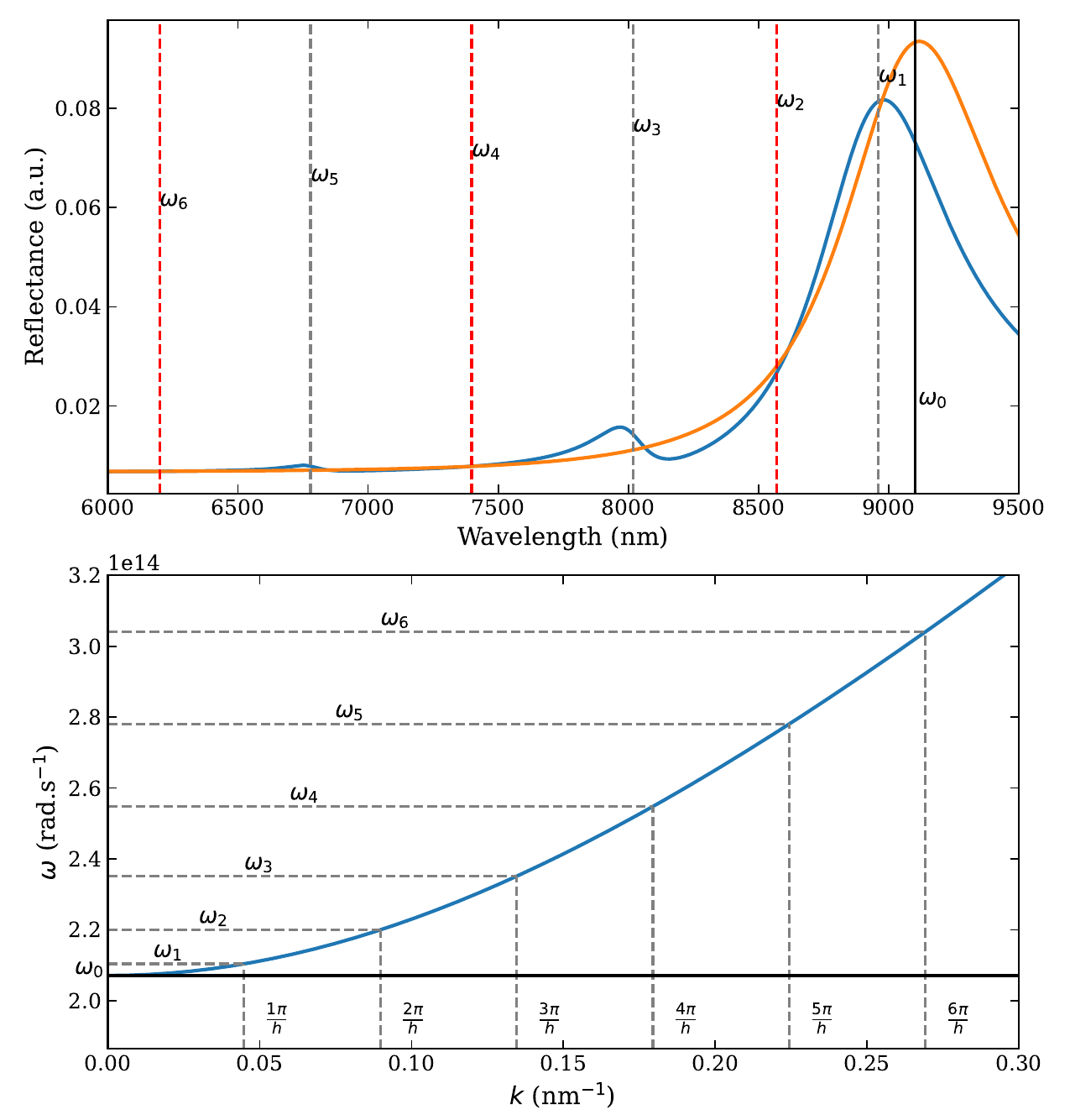}}
\caption{Top: Simulated reflectance spectrum for a 70 nm thick slab without any imaginary part for $\beta^2$ (blue curve) and prediction using the Drude model only (orange curve). Only the odd resonances (grey dashed lines) are visible, while even resonances are absent. A simple cavity formula provides an accurate way to compute the position of these resonances. Bottom: Dispersion curve for plasmons, and cavity resonance conditions $k = \frac{m\pi}{h}$ allowing to compute the resonance frequencies shown at the top of the figure.}
\label{fig:final}
\end{figure}

Despite its simplicity and the fact that the losses are neglected, this formula is quite accurate, as shown in Fig.~\ref{fig:final}, where the lossless dispersion relation is used to compute the position of the resonance on the spectrum. The spectrum shown is computed with $\xi=0$ to better see the position of the resonances. The figure shows clearly that only the resonances corresponding to an odd value of $\ell$ can be seen, as already noticed by Melnyk and Harrisson\cite{melnyk1968resonant} who did not provide any explanation, though.  

The cavity picture allows us to understand why the first resonance (the Ferrel resonance) can in fact not occur at precisely $\omega_0$. As shown in Figure \ref{fig:final}, there is a shift between the position of the first resonance predicted by the hydrodynamic model and by a Drude model, which assumes a rigid electron gas. While this shift is small for such a thick structure, it becomes much larger for thinner structures. Such a large shift is thus a clear sign of a nonlocal response, dictated by the dispersion relation of plasmons. It gives a clear indication on the value of $\beta$. This simple model also allows us to understand that, when $\beta$ tends to zero, all the resonances coalesce into a single resonance, corresponding to the Ferrel resonance within the framework of the Drude model.

The fact that the even resonances of a cavity cannot be excited is a phenomenon that typically arises when a cavity is excited in phase and with the same amplitude from both sides, a phenomenon sometimes referred to as coherent perfect absorption \cite{baranov2017coherent} or more accurately as the interferometric enhancement of the absorption \cite{lemaitre2017interferometric}.

At first glance, the physical situation here does not seem to fit this description because the structure is illuminated from above only. However, the slab functions as a two-mode cavity\cite{pitelet2017fresnel}. The transverse wave also traverses the semiconductor slab, and since we are very close to $\omega_0$, the effective refractive index approaches zero, resulting in a vanishing wavevector and a diverging wavelength. To be precise, at the considered incidence angle, the transverse wave is usually evanescent in the medium but has a very large penetration depth. In both scenarios, the slab thickness is small compared to either the wavelength or the penetration depth, allowing it to reach the bottom interface while maintaining the same phase and amplitude. The plasmon that is generated at the bottom interface is then expected to have the same phase as the plasmon generated at the top interface and a comparable amplitude, thus canceling the even resonances of the cavity. 

We underline that this also leads to the enhancement of odd resonances. This is fundamental in the so-called coherent perfect absorption and explains the large absorption cross-section of plasmonic nanocavities\cite{moreau2012controlled} because it enhances absorption at resonance by a factor of typically four. Without such a phenomenon, it is likely only the main resonance could have been observed here.

\begin{figure}[h!]
    \centering
    \includegraphics[width=\linewidth]{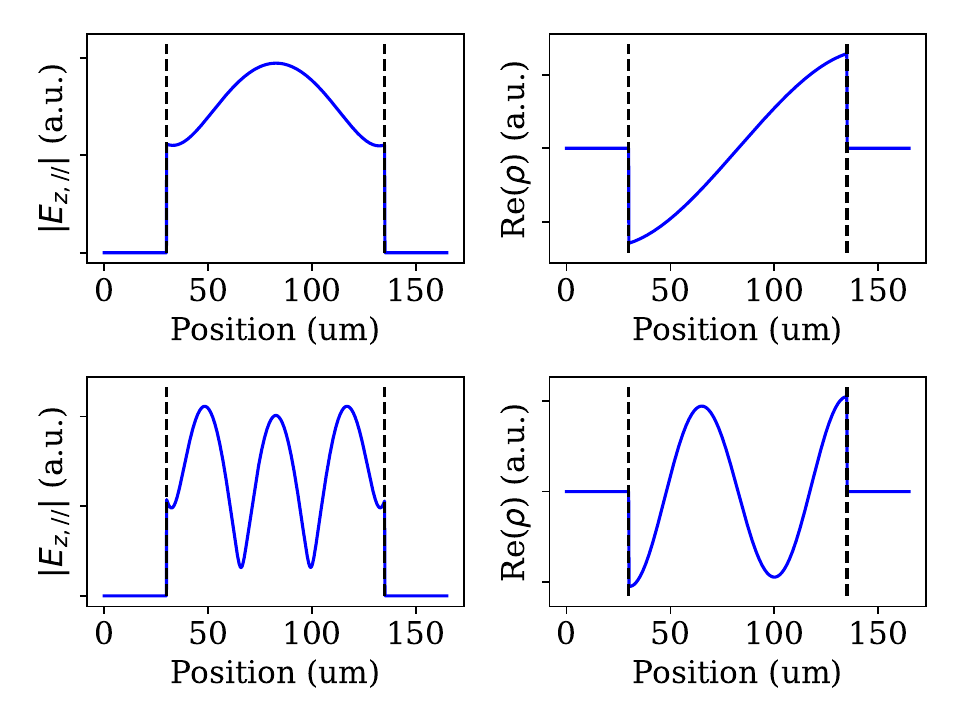}
\caption{Field profiles confirming selective excitation of odd plasmon modes. Top: Modulus of $E_z$ and real part of the charge density for th fundamental mode. Bottom: Same quantities for the mode characterized by $\ell=3$. The antisymmetric charge distribution for the fundamental mode explains the restoring force responsible for the main resonance.}    \label{fig:fields}
\end{figure}

To verify the cavity picture, we extended the PyMoosh simulation library\cite{langevin2024pymoosh} to handle nonlocal response using a generalized S-matrix formalism\cite{benedicto2015numerical} and compute charge density distributions within the semiconductor. Figure \ref{fig:fields} displays the spatial profiles at resonance frequencies, showing the characteristic odd symmetry of the excited modes. The charge density extends deep into the bulk rather than being confined to interfaces as in the Drude model, confirming the bulk plasmon nature of these resonances. This visualization directly demonstrates why even modes remain dark: the symmetric excitation from both interfaces naturally selects odd-symmetry modes.

\section{Experimental results}

\subsection{Sample preparation}

We prepared two series of samples of Si doped n-InAsSb, which we will denote \#1 and \#2. The two series differ by their doping level. Each series derives from a single 204 nm thick n-InAsSb which is subsequently etched to guarantee both an homogeneous level of doping in each series and a good control of the thickness. 

The initial samples were grown using a RIBER-C21 solid source molecular beam epitaxy tool on n-doped ($1-2\times 10^{18}$ cm$^{-3}$) GaSb substrates. Prior to the growth process of the n-InAsSb layer, a thermal desorption step is performed to remove the native oxide and a 210 nm thick buffer layer of undoped GaSb is grown to smooth the surface and bury any impurities remaining after deoxidation.

The n-InAsSb layer is a digital alloy, i.e. a short period superlattice, which allows the deposited thickness, 204 nm, to be accurately monitored for both samples. The doping level of the Si doped n-InAsSb layer, estimated using the optical method described in \cite{taliercio2014brewster},  is of $5.3\times 10^{18}$ cm$^{-3}$ and $1.20\times 10^{19}$ cm$^{-3}$ for sample \#1 and \#2, respectively. 
 While surface accumulation layers are well-documented in InAs-based materials \cite{wieder1974transport,casias2019carrier,rogalski2020inassb}, with typical densities of $\sim 10^{12}$ cm$^{-2}$ (corresponding to $\sim 10^{18}$ cm$^{-3}$ over a few nm), our intentional doping levels are significantly higher. At these high bulk doping concentrations, any surface effect would manifest as charge depletion rather than accumulation, which would shift the plasma frequency to longer wavelengths—opposite to the spectral features observed experimentally. Therefore, we can reliably consider the doped layer as homogeneous for our optical analysis.
The determination of the doping level, as mentioned above, is a reflection experiment under polarized light with an angle of incidence of 60°. A home-made abacus then allows to deduce the carrier density from the measurement of the resonance wavelength\cite{taliercio2014brewster}. 

The initial samples are broken into several pieces and each one is then etched to obtain a series of samples, with a thickness varying from 7 nm to 204 nm. The samples are wet etched at 20°C in a 2:1 ratio of citric acid ($C_6H_8O_7$) and hydrogen peroxide ($H_2O_2$) solution, the time of exposition allowing to control the final thickness. To evaluate the final thickness of each sample, a part of the surface is protected from wet etching by a small drop of photoresist (AZMIR-701) deposited on one edge and heated at 90°C for 1 min 30 s. After removal of the photoresist, the remaining step is measured by atomic force microscopy (AFM) and subtracted from the total thickness of 204 nm. The resulting thicknesses, despite similar exposure time, differ for series \#1 and \#2. 

\subsection{Optical response measurement}

\begin{figure}[h!]
\centering
{\includegraphics[width=\linewidth]{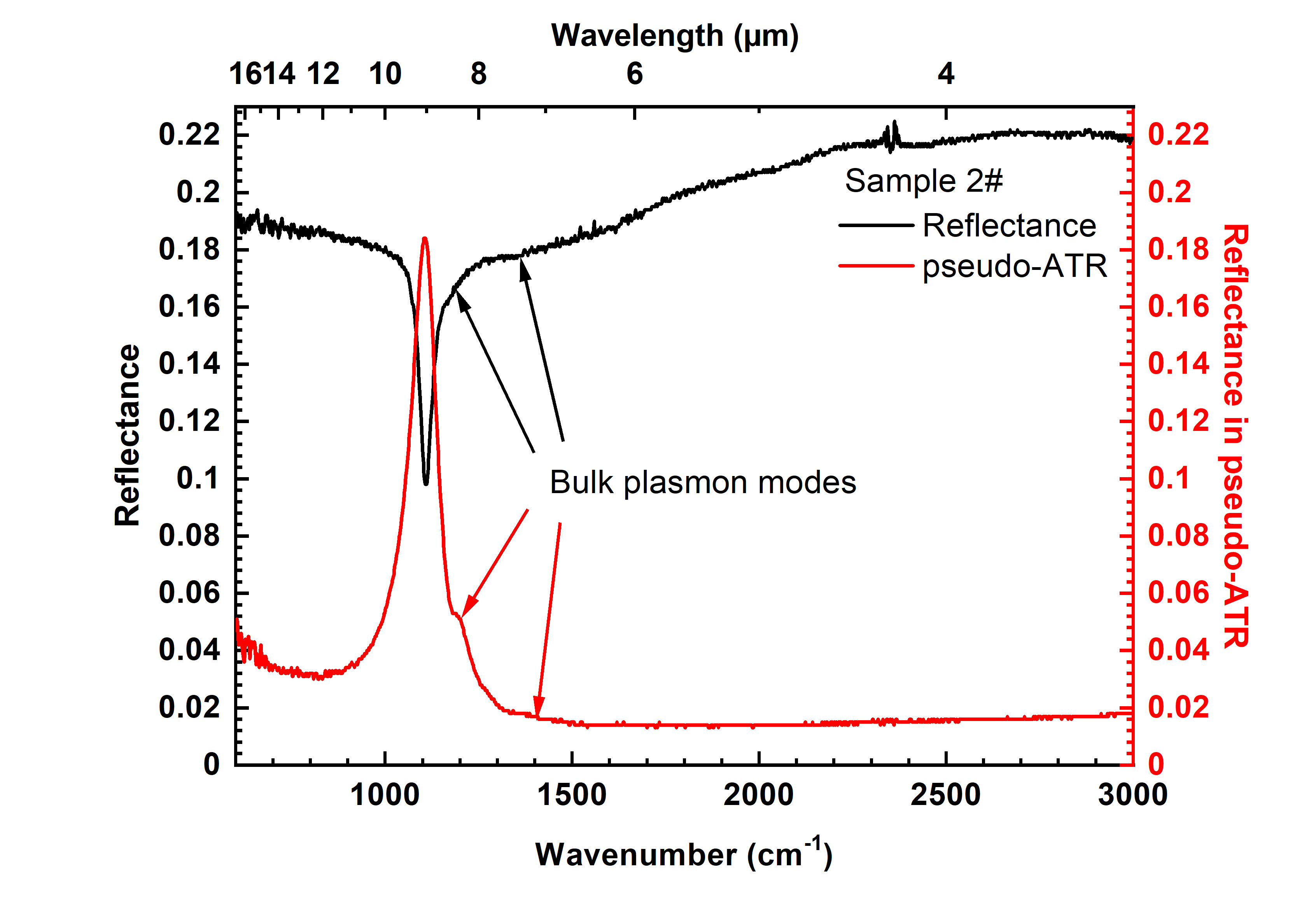}}
\caption{Reflectance spectrum under p-polarized light with an angle of incidence of 60° (black line) and pseudo-ATR spectrum of sample \#1 with a thickness of 204 nm.}
\label{fig:Brewster}
\end{figure}

A typical reflectance spectrum is shown in Figure \ref{fig:Brewster} (black line) for sample \#2. 
The dip in the reflectance that can be observed around 9 µm corresponds to the excitation of the main plasmon resonance inside the InAsSb layer. 

\begin{figure}[h!]
\centering
{\includegraphics[width=0.9\linewidth]{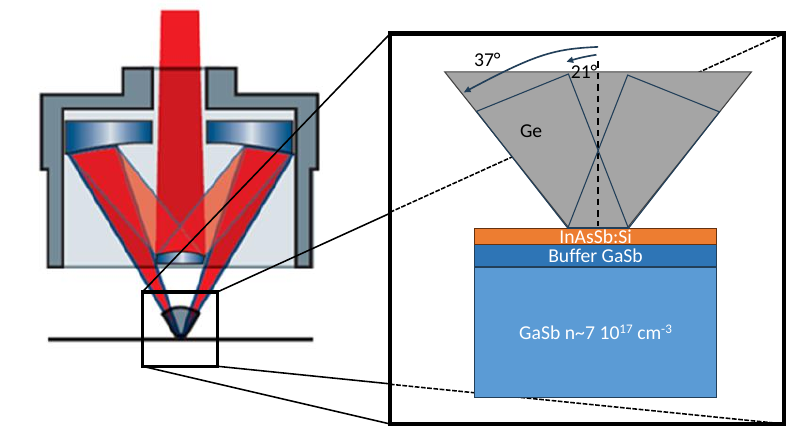}}
\caption{Experimental setup in the pseudo-ATR configuration. The sample is illuminated with incidence angle ranging from 21$^\circ$ to 37$^\circ$. The Ge prism is in direct contact with the sample, but given the incidence angles, no total internal reflection occurs.}
\label{fig:setup}
\end{figure}

This resonance can be excited either in reflection, when illuminated from air, or when illuminated from inside a germanium crystal using a Schwarzschild objective, mounted on a microscope (Hyperion 3000) and coupled to an FTIR spectrometer (Bruker, Vertex 70). Given the high index of the Ge crystal (around 4), such a configuration amounts to exciting the structure using a prism, with an angle of incidence between 21.5° and 37°, determined by the objective (see Fig.\ref{fig:setup}). For media with a refractive index smaller than 1.6, this means total internal reflection will occur and this would constitute an attenuated total internal reflection (ATR) setup. However, especially in the mid-IR range, materials most often present larger indexes so that we call this configuration "pseudo-ATR" in the rest of the paper. 

The pseudo-ATR spectrum of sample \#2 is the red curve in Figure \ref{fig:Brewster}, showing that the fundamental plasmon resonance leads to a reflectance peak. Furthermore, this allows to observe supplementary resonances that are more easily spotted in pseudo-ATR configuration but that can also be found in the reflectance when illuminating the structure from air, when the spectrum is very closely analyzed. This underlines one of the main contribution of our work: when the doped semi-conductor is sandwiched between two materials (Ge and GaSb) with large and close refractive index, plasmon resonances are much easier to observe than in any other configuration. 

\subsection{Plasmon resonances\label{sec:results}}

Using samples of different thicknesses, we systematically map the plasmon resonances visible in our pseudo-ATR configuration. The enhanced visibility afforded by this setup allows us to observe resonances up to order $\ell=11$ for sample \#2, far exceeding previous reports\cite{de2018viscoelastic}. While these high-order resonances appear at longer wavelengths due to our thicker samples and lower doping levels, they demonstrate that bulk plasmon physics is not confined to ultra-thin films but persists in semiconductors nearly an order of magnitude thicker.

\begin{figure}[h!]
\centering
{\includegraphics[width=\linewidth]{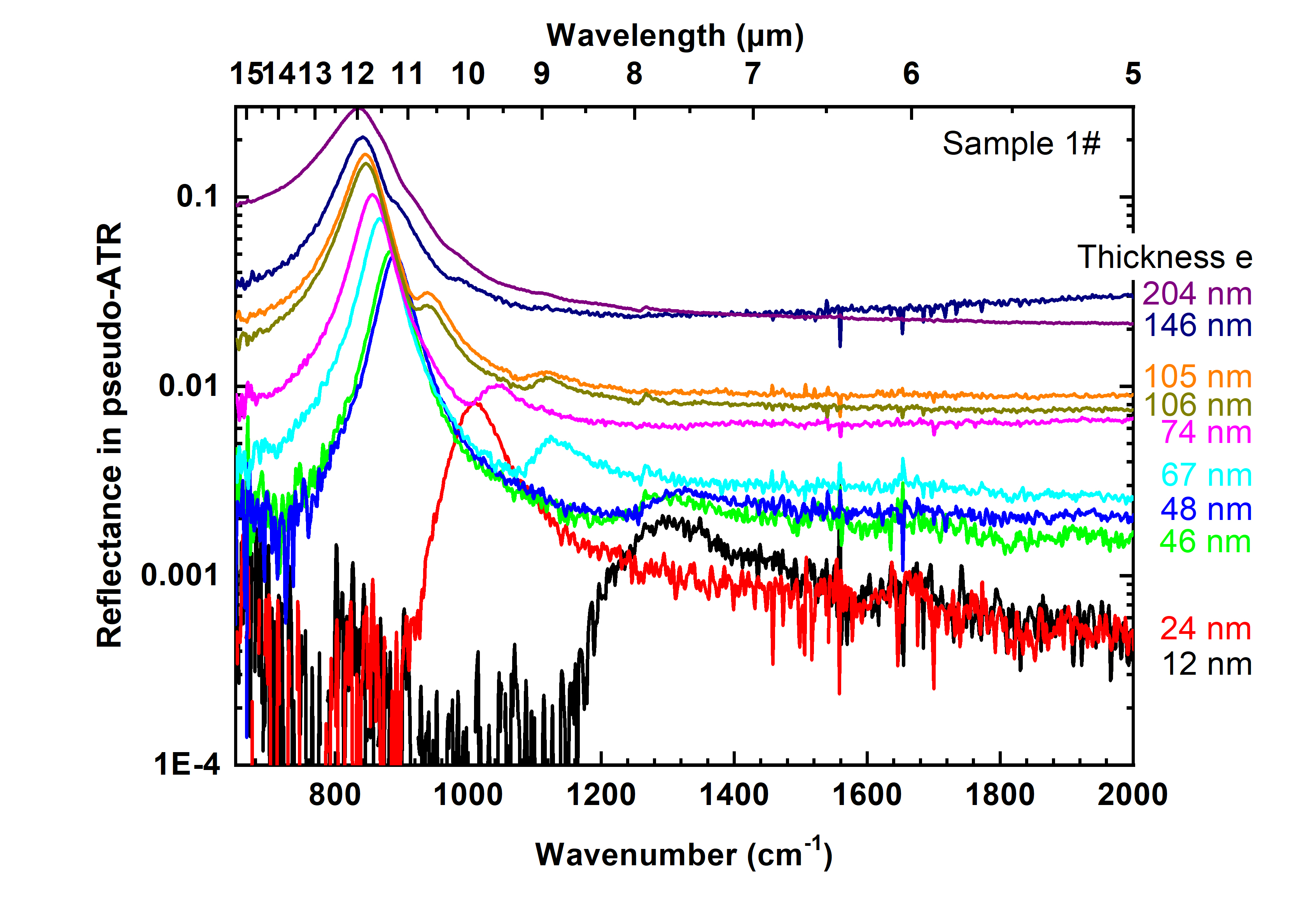}}
\caption{Pseudo-ATR reflectance spectra for several thickness of sample \#1.}
\label{fig:ATR1}
\end{figure}

The pseudo-ATR spectra of series \#1 are shown in Figure \ref{fig:ATR1} in log scale. Spectra are background corrected with a reference sample of GaSb with a 210 nm thick buffer layer of undoped GaSb. In each spectrum the Ferrel resonance can be seen at short wavelength and it progressively blueshifts when the thickness decreases. This blueshift cannot be explained within the Drude model where the Ferrell resonance position is thickness-independent, providing clear evidence of nonlocal effects described by the plasmon dispersion relation.

We identify this peak as the first order plasmon mode, $\ell = 1$. At higher wavelengths, some additional peaks corresponding to the odd high order plasmon modes, $\ell = 3, 5, 7, ...$ can be observed, even if higher order resonances can be difficult to identify. All these resonances also blueshift with decreasing thickness. Note that the small peak at 1250 cm$^{-1}$ is due to the GaSb substrate. To accurately extract the wavenumber of each high-order plasmon resonance, we first fit the first-order plasmon resonance with a Lorentzian function, which we subtract from the experimental data. This enhances the visibility of the high order resonances.

\begin{figure}[h!]
\centering
{\includegraphics[width=\linewidth]{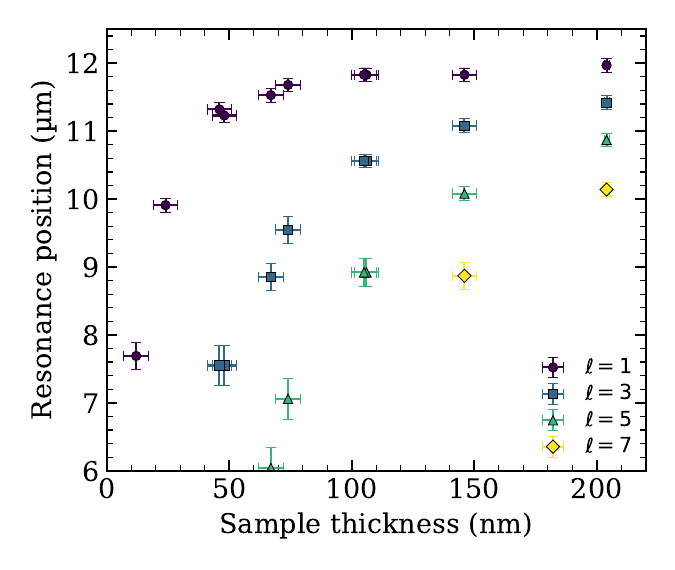}}
\caption{Position of the plasmon resonances as determined using the experimental data for sample \#1. We identify each resonance with an odd number. Error bars are the vertical and horizontal ticks for each disk.}
\label{fig:logATR1}
\end{figure}

The position of the resonances as a function of the sample thickness is summarized in Figure \ref{fig:logATR1}. The colored disks correspond to the wavelength of the plasmon resonances, white, red, blue and pink for $\ell = 1, 3, 5$ and $7$ respectively. Error bars on the disk position are indicated with vertical and horizontal lines.

All plasmon resonances clearly blueshift with decreasing thickness, following the cavity resonance condition $k = \ell\pi/h$ combined with the plasmon dispersion relation, as predicted by Eq.~\ref{eq:cavity}.

\begin{figure}[h!]
\centering
\includegraphics[width=\linewidth]{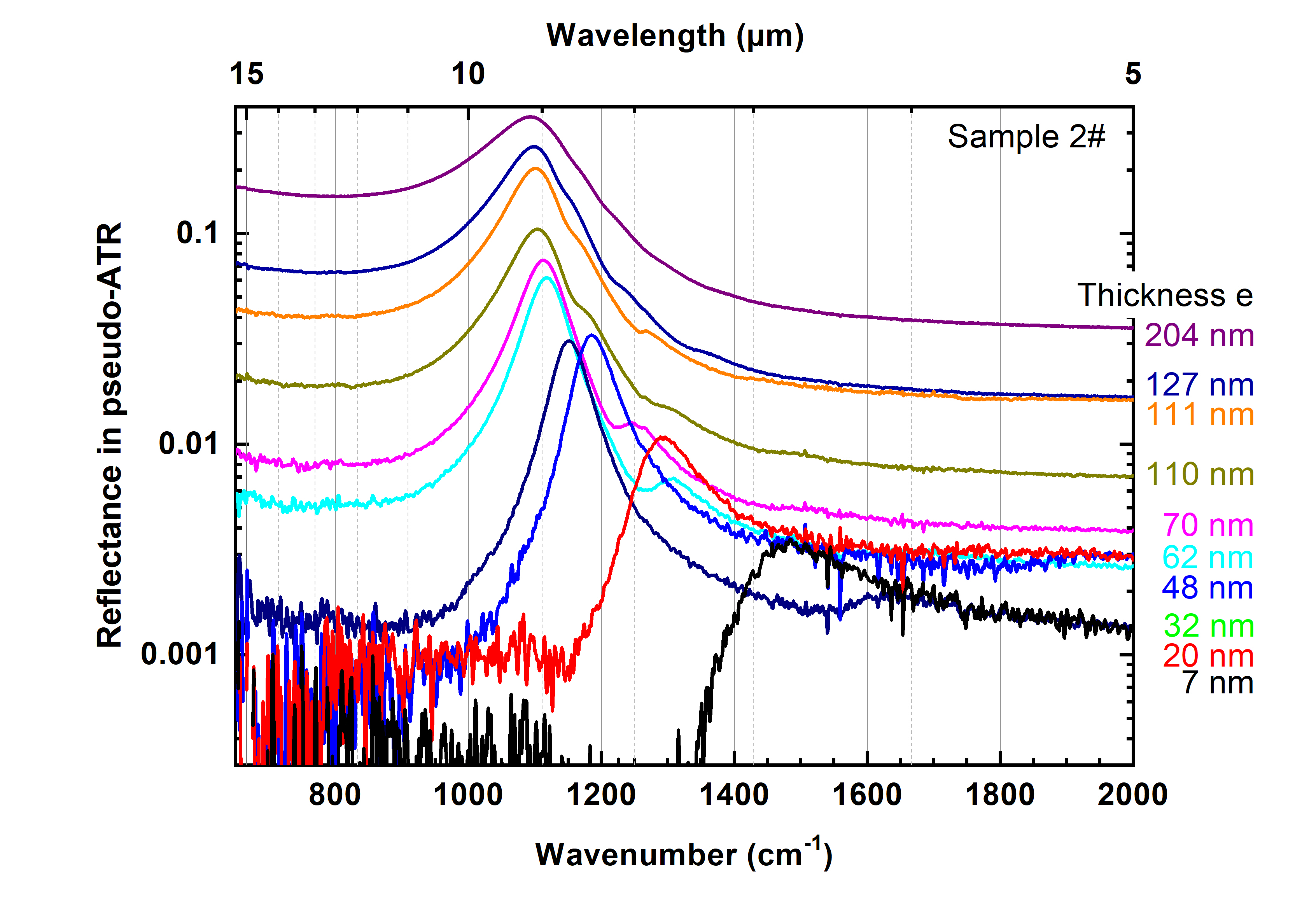}
\caption{Pseudo-ATR spectra for several thicknesses, for sample series \#2.}
\label{fig:ATR2}
\end{figure}

The same study is carried out on sample \#2. Figure \ref{fig:ATR2} is the pseudo-ATR spectra for different thicknesses. As the sample presents a higher level of doping, the first order of the bulk plasmon mode is blue-shifted compared to sample \#1. The resonances are also broadened. However, it is possible to observe the high order bulk plasmon modes up to $\ell = 11$. To our knowledge, this represents the highest-order bulk plasmon resonance reported in doped semiconductors, made possible by the symmetric refractive index configuration and the relatively low losses in our samples.


\begin{figure}[h!]
\centering
{\includegraphics[width=\linewidth]{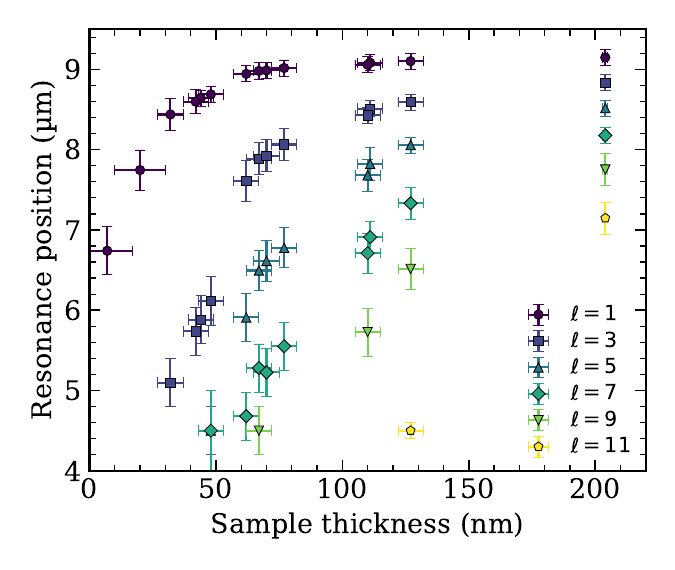}}
\caption{Position of the plasmon resonances as determined using the experimental data for sample \#2. We identify each resonance with an odd number. Error bars are the vertical and horizontal ticks for each disk.}
\label{fig:logATR2}
\end{figure}

The data provide two key signatures for parameter extraction: (i) the blueshift of resonances away from $\omega_0$, which directly determines Re($\beta$) via Eq.~\ref{eq:cavity}, and (ii) the absence of the sharp Fano profiles predicted for real $\beta$ (cf. Fig.~\ref{fig:final}), requiring an imaginary part to explain the observed smoothing. These features are sufficiently distinctive that even a single thin sample with visible high-order resonances contains enough information to extract all the model parameters as demonstrated in the following section.

\section{Improved parameter retrieval}

To streamline the parameter retrieval process, we have developed an automated method to retrieve these parameters. Our method includes three key components. Firstly, we use the hydrodynamic model with a complex hydrodynamic parameter $\beta$, which will be explained in detail below. Secondly, we define a cost function that prioritizes matching the overall shape of the curves rather than just minimizing the distance between data points, aligning better with the intuitive assessment of physicists. Thirdly, we employ the Differential Evolution (DE) global algorithm, known for its efficiency in solving physical problems\cite{barry2020evolutionary,bennet2024illustrated}, ensuring that we do not miss any satisfactory parameter values.

\begin{figure}[h!]
\centering
{\includegraphics[width=\linewidth]{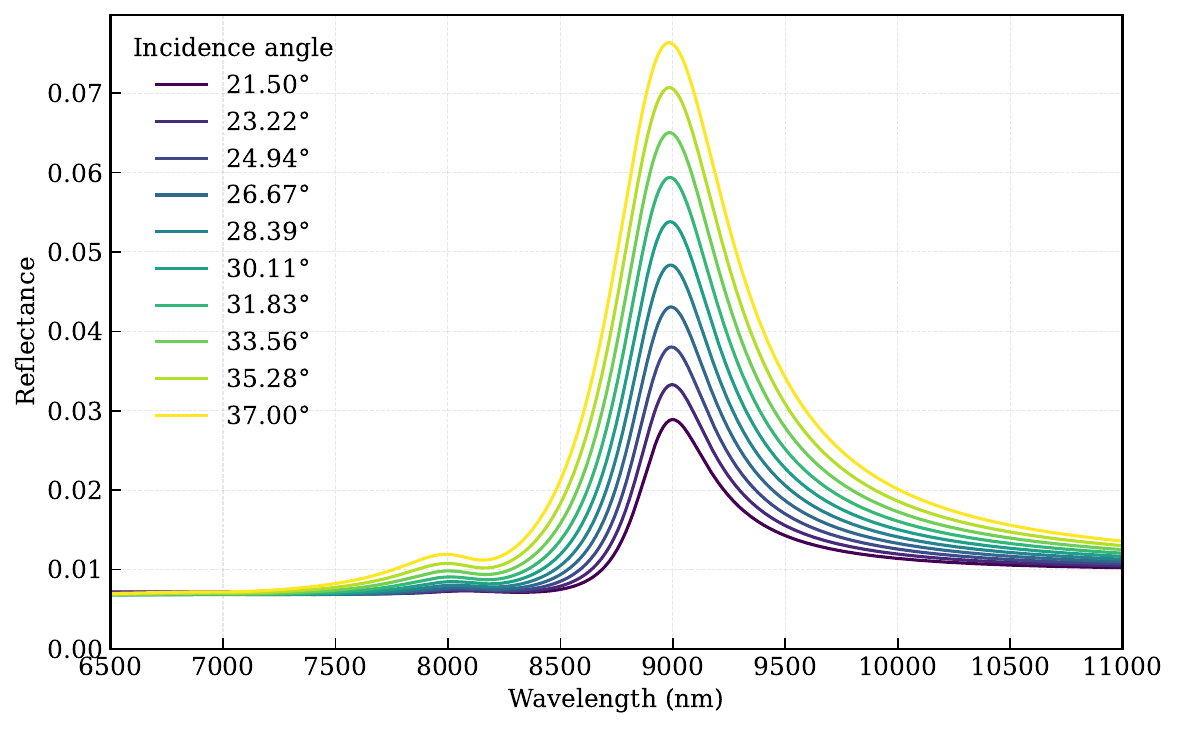}}
\caption{Simulations showing the dependence of the reflectance spectrum with respect to the incidence angle for angles ranging from 21.5° to 37° for a 70 nm thick InAsSb.}
\label{fig:angle}
\end{figure}

The experimental results in the pseudo-ATR configuration cover a wide range of incidence angles, from 21.5° to 37°, with no means to determine the respective weight of each angle. However, as shown in Figure \ref{fig:angle}, the response of the semiconductor slab strengthens and the visibility of the plasmon resonances improves with increasing incidence angle. We also note that the overall shape of the signal remains fairly consistent. Therefore, we assume that the spectrum corresponding to the largest angle likely dominates the experimental results. We thus fit the data to the simulation results using only the largest incidence angle, but we do not expect a good match below $\omega_0$ since the reflectance for the lower incidence angle is nearly the same as for the highest. We thus expect the single-angle simulation to underestimate the reflectance for large wavelengths.

We first chose to fit the experimental data using parameters that rely minimally on underlying assumptions and can be considered empirical, such as $\chi_b$, $\omega_p$, and $\gamma$ for the Drude model, and a complex value of $\eta^2$ for the hydrodynamic part:
\begin{equation}
    \eta^2 = \beta^2 - i\,\omega \frac{\xi}{n_0 m^*}.
\end{equation}
We actually introduce $\tau = \frac{\xi}{n_0 m^*}$ as a parameter to be optimized and retrieve $\xi$ afterwards. Finally, we also include a parameter for the background signal and another for scaling the data.

The cost function we chose is a linear combination of (i) the difference between the model and the experimental data, as is usual, and (ii) the difference between the slopes of the model and the experimental data, normalized. Including the difference in slopes as a criterion results in a model curve that may be shifted but maintains the same overall shape, aligning with what a physicist would intuitively aim for.

\begin{widetext}
The cost function is given by 
\begin{equation}
f(\Pi) = \frac{1}{N} \sum_{i=1} 
\left| R(\lambda_i) - S(\Pi,\lambda_i) \right|
+ N \times 
\frac{1}{N-1}\sum_{i=1}^{N-1} 
\left|
(R(\lambda_{i+1})-R(\lambda_i))
-(S(\Pi,\lambda_{i+1})-S(\Pi,\lambda_i))
\right| 
\end{equation}
where $\lambda_i$ is a wavelength for which a reflectance $R(\lambda_i)$ has been measured, $\Pi$ represents the parameters of the simulation ($\chi_b$, $\omega_p$, $\gamma$, $\beta$, $\tau = \frac{\xi}{n_0\,m^*}$, background signal intensity, scaling parameter), and $S(\Pi,\lambda_i)$ is the reflectance of the structure simulated for the given parameters and wavelength.
\end{widetext}

The second term in the cost function guides the optimization to minimize differences in the derivatives of the experimental and simulated reflectance curves. While the experimental signal $R(\lambda)$ naturally includes noise that increases the cost function’s value, this noise does not alter the position of the minimum in the parameter space, ensuring the robustness of the optimization process.

In order to find the minimum of the cost function, we use Differential Evolution (DE) (more precisely its Quasi-Oppositional version and then a steepest descent to refine the results\cite{bennet2024illustrated}) to look for satisfactory values of the parameters.

The numerical computation of the reflectance, taking into account an imaginary part for $\beta^2$, has been included in the open source PyMoosh software\cite{langevin2024pymoosh}. In addition, we have made the code we used available under the form of a Jupyter notebook, we is made possible because PyMoosh combines simulation and optimization tools\cite{latest}.

\begin{figure}[h!]
\centering
{\includegraphics[width=\linewidth]{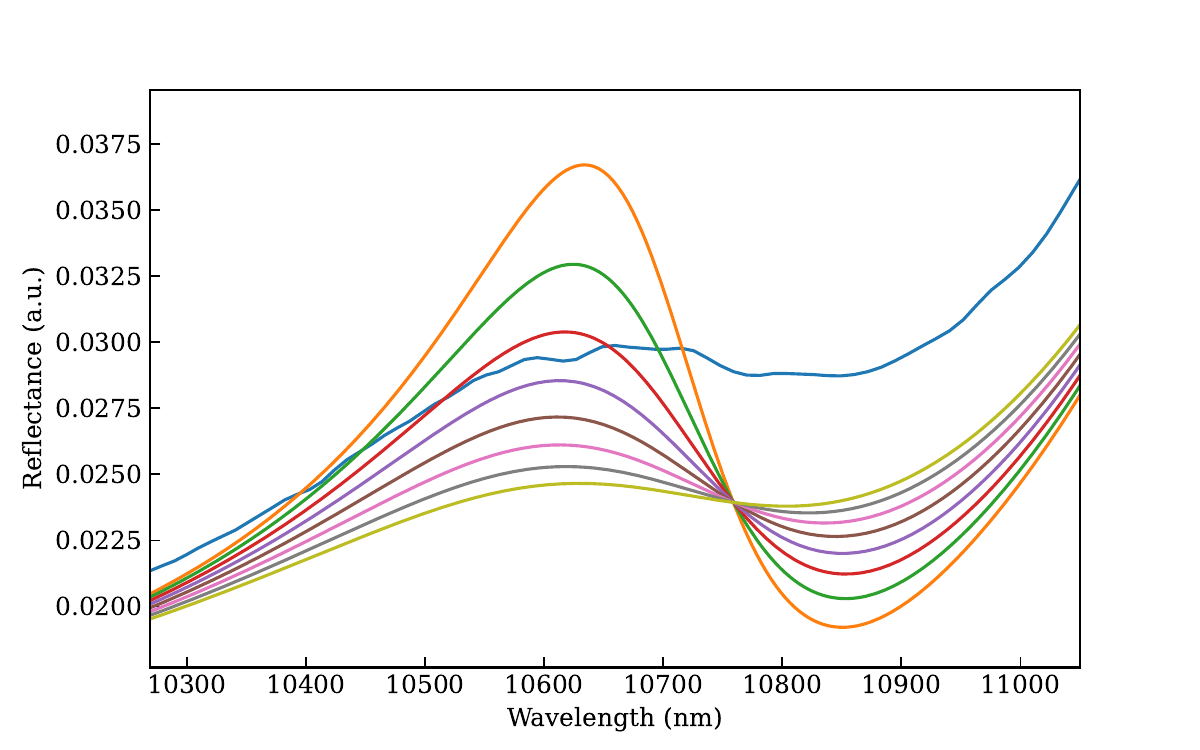}}
\caption{Resonance characterized by $\ell=3$ for a 106 nm thick sample of series \#1, for different values of  $\frac{\xi}{n_0 m^*}$ ranging from 0 to 7$\times 10^{-4}$ in m$^2$.s$^{-1}$. The orange curve corresponds to $\xi = 0$. When the value of $\xi$ increases, the resonance profile becomes smoother and deviates from a Fano profile.}
\label{fig:losses}
\end{figure}

The importance of the imaginary part of the $\beta$ parameter is illustrated in Figure \ref{fig:losses}. As the imaginary part increases from zero to $8\times 10^{13}$ m$^2$.s$^{-2}$, the resonance profile changes drastically. 
When $\beta$ is real, the resonance exhibits a clear Fano profile, indicating strong coupling of the slab to the continuum on both sides -- which can be linked to our experimental setup. However, as the imaginary part grows, the profile becomes smoother. This significant change in the profile allows for a rather accurate estimation of the imaginary part, for which a value of zero can be completely excluded.

\begin{figure}[h!]
\centering{\includegraphics[width=\linewidth]{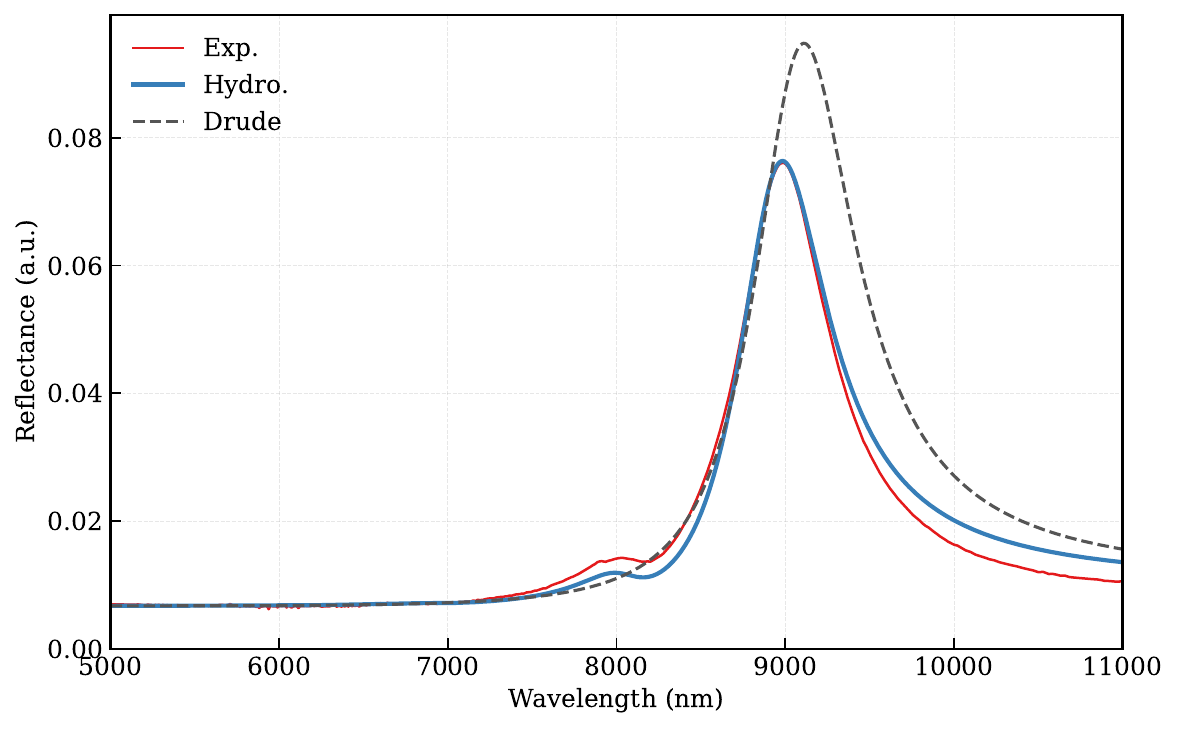}}
\caption{Experimental reflectance data for a 70 nm thick InAsSb layer of sample \#2 in the pseudo-ATR configuration, compared with the best fit obtained for the given model parameters. The predictions of the Drude model with the same parameters are also shown, to illustrate the necessity of the nonlocal description.}
\label{fig:monofit}
\end{figure}

The parameter retrieval can be performed on a single spectrum for a given thickness or across multiple thicknesses. Based on our experience, a fit on a single spectrum is sufficient when the main resonance and at least one higher-order resonance are clearly visible. The more resonances that are observable, the better the determination of $\beta$. For sample \#2, selecting a spectrum with enough resonances yielded the most reliable results. For instance, using the spectrum for a 100 nm thickness (shown in Fig. \ref{fig:monofit}), we obtained estimations of $n_0 =1.40 \pm 0.08 \times 10^{19}$ cm$^{-3}$, $m^*=0.083\pm0.006\,m_0$ and $\xi = 3.7\pm 1.0\times 10^{-10}$ Pa.s.

\begin{figure}[h!]
\centering
{\includegraphics[width=\linewidth]{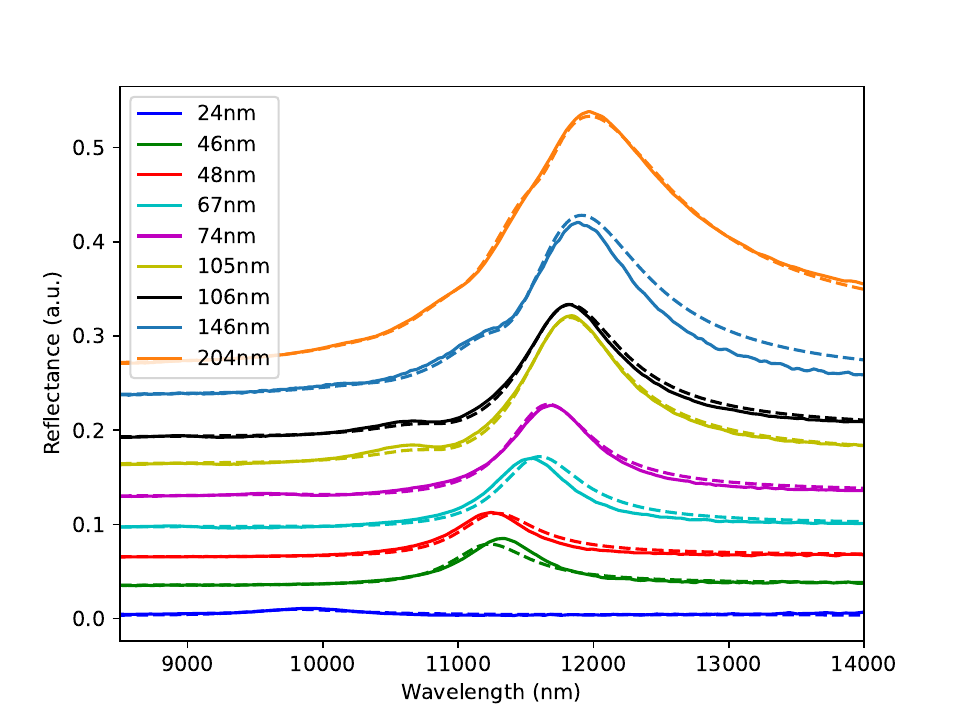}}
\caption{Experimental reflectance data (solid lines) for several thicknesses and the result of a global fit (dashed lines) on all the curves at once for sample \#1. The curves are shifted vertically to be more clearly visible.}
\label{fig:grosfit}
\end{figure}

However, to get more reliable values, a fit taking into account all the thicknesses can be preferable. At small thicknesses, the shift of the main resonance away from $\omega_0$ is directly linked to the value of $\beta$, thus providing important information for the optimization. For sample \#1, the results are very convincing, as shown Fig. \ref{fig:grosfit}. The values obtained are $n_0 = 5.31 \pm 0.5\times 10^{18}$ cm$^3$, $m^* = 0.057\pm0.05 \,m_0$ and $\xi = 2.11\pm 0.6\times10^{-10}$ Pa.s. 

Uncertainty quantification for this inverse problem presents unique challenges that standard methods do not fully address. While multiple optimization runs consistently converge to similar parameter values (with a ~5\% variability typically), suggesting robust convergence despite the presence of multiple local minima, this statistical variability underestimates the true parameter uncertainty. 

We therefore adopted a physics-informed sensitivity analysis: parameters were varied from their optimal values until the fit quality degraded beyond what would be considered acceptable. This approach yields conservative uncertainty estimates of approximately 10\% for $n_0$ and $m^*$, capturing not only numerical variability but also the inherent ambiguity in defining acceptable fit quality given experimental noise and systematic effects. The uncertainty on the bulk viscosity $\xi$ is larger, as the resonance profile shows a gradual dependence on this parameter (see Fig. \ref{fig:losses}), making precise determination more challenging.

While rigorous uncertainty estimation for such inverse problems remains an open challenge, we anticipate that improvements to the experimental setup—particularly better control of the incidence angle distribution and more accurate thickness measurements—would significantly reduce these uncertainties before more sophisticated analysis methods become necessary. Indeed, the current approach of fitting multi-angle experimental data with plane-wave simulations already introduces systematic approximations that likely dominate over numerical uncertainties.

Beyond the technical aspects of parameter retrieval, our results raise fundamental questions about the nature of electron transport in semiconductors. The observation of significant bulk viscosity in electron gases is surprising. Stokes' hypothesis —that bulk viscosity can be neglected— is widely adopted in fluid dynamics, and theoretical studies suggest it should be particularly valid for monoatomic gases lacking internal degrees of freedom\cite{graves1999bulk}. The electron gas, being fundamentally a collection of identical particles, might be expected to behave similarly\cite{ciraci2017current,conti1999elasticity}. We suspect the available states in the conduction band play a role analogous to the internal degrees of freedom in polyatomic gases, though the microscopic theory connecting band structure to bulk viscosity in electron gases remains to be developed.

The GNOR model and shear viscosity approaches lead to incompatible frequency dependencies for the nonlocal response\cite{de2018viscoelastic}. While our results appear to favor the GNOR framework through the identification of a bulk viscosity term, the full picture remains incomplete. Notably, we observe systematic discrepancies in higher-order resonances: their profiles cannot be accurately reproduced with a constant $\xi$. This suggests a frequency-dependent bulk viscosity $\xi(\omega)$ that would decrease with increasing frequency. This suggested decrease of $\xi$ with frequency aligns with the typical behavior of bulk viscosity in complex fluids, where high-frequency oscillations progressively decouple from slower relaxation processes. This points toward the need for models that go beyond simple constant coefficients, though more controlled experiments — particularly with better-defined incidence angles and precise sample thickness measurements — will be essential to fully characterize this dispersion\cite{ciraci2017current} and advance our theoretical understanding of nonlocal effects in electron gases.

\section{Conclusion and perspectives}

We have proposed a setup that maximizes the visibility of plasmon resonances in highly doped semiconductors. Combined with the numerical methods detailed in this work, our findings show that the hydrodynamic model, with a complex nonlocal parameter $\beta^2$, is remarkably sufficient to capture the optical response of these materials with a high level of accuracy. 
We have also identified a physically consistent origin for the imaginary part of $\beta^2$, grounded in the bulk viscosity (or second viscosity) of the electron gas, offering a convincing and straightforward justification for the well-known GNOR model\cite{mortensen2014generalized,Raza2015,boroviks2022extremely}.

We have also introduced a retrieval technique based on an adapted cost function and the use of a global optimization algorithm. In the theoretical framework we propose, this technique could allow for the estimation of the doping level, the effective mass of electrons, the repulsion between electrons and even the bulk viscosity of the electron gas with a single optical measurement - without any abacus. Additionally, we have shown that these parameters can be retrieved manually with satisfactory accuracy, provided that the resonances are clearly identified, making the approach accessible even without advanced computational tools.

While plasmon resonances have a strong fundamental interest, it is also crucial to highlight that spatial dispersion plays an important but distinct role below the epsilon-near-zero (ENZ) frequency. In this frequency range, highly doped semiconductors emerge as exceptional plasmonic materials capable of supporting surface waves or more complex guided modes with high effective indices (slow light), which can be exploited to design sensors for detecting and characterizing even small quantities of molecules\cite{smaali2021reshaping,paggi2023over}. Previous studies have demonstrated that the slower the plasmonic guided mode\cite{ajib2019energy}, the more pronounced the impact of spatial dispersion\cite{Moreau2013,raza2013nonlocal,khalid2021influence,boroviks2022extremely}. This effect is particularly significant near the ENZ frequency and in the presence of high refractive index materials\cite{pitelet2019influence}, as commonly observed in the IR range where semiconductors exhibit large permittivities. Therefore, we anticipate that nonlocal effects will remain significant even below the ENZ frequency, underscoring the importance of having suitable simulation tools and parameter retrieval techniques.

Despite the challenges in observing nonlocal effects in metals within the visible range, significant efforts have been made over the past decade to develop theoretical and numerical tools\cite{Toscano2012,schmitt2016dgtd,kwiecien2023nonlocal} and to design structures that maximize these effects. It is striking how these advancements are now proving to be highly relevant for highly doped semiconductors. Given the crucial role of spatial dispersion in these materials, as discussed above, it seems likely that spatial dispersion will soon become a standard approach for describing their optical response. We are pleased to see this area of research transition from a purely theoretical curiosity to a practical, widely-adopted tool and are convinced that this shift will help to answer the remaining fundamental questions.

\section*{Acknowledgments}

This work has been partially funded by the French “Investment for the Future” program (EquipEx EXTRA, ANR 11-EQPX-0016), and by the French ANR (Nano-Elastir, 19-ASTR-0003-02; SWAG-P 23-CE09-0014).

\bibliography{references}

@article{buresti2015note,
  title={A note on Stokes’ hypothesis},
  author={Buresti, Guido},
  journal={Acta Mechanica},
  volume={226},
  pages={3555--3559},
  year={2015},
  publisher={Springer}
}

@article{bandurin2016negative,
  title={Negative local resistance caused by viscous electron backflow in graphene},
  author={Bandurin, DA and Torre, Iacopo and Kumar, R Krishna and Ben Shalom, M and Tomadin, Andrea and Principi, A and Auton, GH and Khestanova, E and Novoselov, KS and Grigorieva, IV and others},
  journal={Science},
  volume={351},
  number={6277},
  pages={1055--1058},
  year={2016},
  publisher={American Association for the Advancement of Science}
}

@article{polini2020viscous,
  title={Viscous electron fluids},
  author={Polini, Marco and Geim, Andre K},
  journal={Physics Today},
  volume={73},
  number={6},
  pages={28--34},
  year={2020},
  publisher={AIP Publishing}
}

@article{steinmann1960experimental,
  title={Experimental verification of radiation of plasma oscillations in thin silver films},
  author={Steinmann, Wulf},
  journal={Physical Review Letters},
  volume={5},
  number={10},
  pages={470},
  year={1960},
  publisher={APS}
}

@article{drude1900elektronentheorie,
  title={Zur elektronentheorie der metalle},
  author={Drude, Paul},
  journal={Annalen der physik},
  volume={306},
  number={3},
  pages={566--613},
  year={1900},
  publisher={Wiley Online Library}
}

@article{moreau2012controlled,
  title={Controlled-reflectance surfaces with film-coupled colloidal nanoantennas},
  author={Moreau, Antoine and Cirac{\`\i}, Cristian and Mock, Jack J and Hill, Ryan T and Wang, Qiang and Wiley, Benjamin J and Chilkoti, Ashutosh and Smith, David R},
  journal={Nature},
  volume={492},
  number={7427},
  pages={86--89},
  year={2012},
  publisher={Nature Publishing Group UK London}
}

@article{boroviks2022extremely,
  title={Extremely confined gap plasmon modes: when nonlocality matters},
  author={Boroviks, Sergejs and Lin, Zhan-Hong and Zenin, Vladimir A and Ziegler, Mario and Dellith, Andrea and Gon{\c{c}}alves, PAD and Wolff, Christian and Bozhevolnyi, Sergey I and Huang, Jer-Shing and Mortensen, N Asger},
  journal={Nature Communications},
  volume={13},
  number={1},
  pages={3105},
  year={2022},
  publisher={Nature Publishing Group UK London}
}

@article{reynaud2018enhancing,
  title={Enhancing Reproducibility and Nonlocal Effects in Film-Coupled Nanoantennas},
  author={Reynaud, Cl{\'e}ment A and Duch{\'e}, David and Le Rouzo, Judika{\"e}l and Nasser, Antoine and Nony, Laurent and Pourcin, Florent and Margeat, Olivier and Ackermann, J{\"o}rg and Berginc, Gerard and Nijhuis, Christian A and others},
  journal={Advanced Optical Materials},
  volume={6},
  number={23},
  pages={1801177},
  year={2018},
  publisher={Wiley Online Library}
}

@article{graves1999bulk,
  title={Bulk viscosity: past to present},
  author={Graves, Rick E and Argrow, Brian M},
  journal={Journal of Thermophysics and Heat Transfer},
  volume={13},
  number={3},
  pages={337--342},
  year={1999}
}

@article{gad1995stokes,
  title={Stokes’ hypothesis for a Newtonian, isotropic fluid},
  author={Gad-el-Hak, Mohamed},
  journal={Journal of Fluids Engineering},
  volume={117},
  number={1},
  pages={3--5},
  year={1995}
}

@techreport{ash1991second,
  title={Second coefficient of viscosity in air},
  author={Ash, Robert L and Zuckerwar, Allan J and Zheng, Zhonquan},
  institution = "NASA",
  number = "No. NASA-CR-187783",
  year={1991}
}

@article{sakat2021generalized,
  title={Generalized electromagnetic theorems for nonlocal plasmonics},
  author={Sakat, Emilie and Moreau, Antoine and Hugonin, Jean-Paul},
  journal={Physical Review B},
  volume={103},
  number={23},
  pages={235422},
  year={2021},
  publisher={APS}
}

@article{pitelet2019influence,
  title={Influence of spatial dispersion on surface plasmons, nanoparticles, and grating couplers},
  author={Pitelet, Armel and Schmitt, Nikolai and Loukrezis, Dimitrios and Scheid, Claire and De Gersem, Herbert and Cirac{\`\i}, Cristian and Centeno, Emmanuel and Moreau, Antoine},
  journal={JOSA B},
  volume={36},
  number={11},
  pages={2989--2999},
  year={2019},
  publisher={Optica Publishing Group}
}

@article{pitelet2017fresnel,
  title={Fresnel coefficients and Fabry-Perot formula for spatially dispersive metallic layers},
  author={Pitelet, Armel and Mallet, {\'E}milien and Centeno, Emmanuel and Moreau, Antoine},
  journal={Physical Review B},
  volume={96},
  number={4},
  pages={041406},
  year={2017},
  publisher={APS}
}

@article{raza2013nonlocal,
  title={Nonlocal response in thin-film waveguides: loss versus nonlocality and breaking of complementarity},
  author={Raza, S{\o}ren and Christensen, Thomas and Wubs, Martijn and Bozhevolnyi, Sergey I and Mortensen, N Asger},
  journal={Physical Review B—Condensed Matter and Materials Physics},
  volume={88},
  number={11},
  pages={115401},
  year={2013},
  publisher={APS}
}

@article{paggi2023over,
  title={Over-coupled resonator for broadband surface enhanced infrared absorption (SEIRA)},
  author={Paggi, Laura and Fabas, Alice and El Ouazzani, Hasnaa and Hugonin, Jean-Paul and Fayard, Nikos and Bardou, Nathalie and Dupuis, Christophe and Greffet, Jean-Jacques and Bouchon, Patrick},
  journal={Nature Communications},
  volume={14},
  number={1},
  pages={4814},
  year={2023},
  publisher={Nature Publishing Group UK London}
}

@article{smaali2021reshaping,
  title={Reshaping plasmonic resonances using epsilon-near-zero materials for enhanced infrared vibrational spectroscopy},
  author={Smaali, Rafik and Taliercio, Thierry and Moreau, Antoine and Centeno, Emmanuel},
  journal={Applied Physics Letters},
  volume={119},
  number={18},
  year={2021},
  publisher={AIP Publishing}
}

@article{langevin2024pymoosh,
  title={PyMoosh: a comprehensive numerical toolkit for computing the optical properties of multilayered structures},
  author={Langevin, Denis and Bennet, Pauline and Khaireh-Walieh, Abdourahman and Wiecha, Peter and Teytaud, Olivier and Moreau, Antoine},
  journal={JOSA B},
  volume={41},
  number={2},
  pages={A67--A78},
  year={2024},
  publisher={Optica Publishing Group}
}

@article{khalid2021influence,
  title={Influence of the electron spill-out and nonlocality on gap plasmons in the limit of vanishing gaps},
  author={Khalid, M and Morandi, O and Hervieux, PA and Manfredi, G and Cirac{\`\i}, C},
  journal={Physical Review B},
  volume={104},
  number={15},
  pages={155435},
  year={2021},
  publisher={APS}
}

@software{latest,
  author       = {Denis Langevin and Antoine Moreau},
  title        = {AnMoreau/PyMoosh: 4.0.1},
  month        = nov,
  year         = 2025,
  publisher    = {Zenodo},
  version      = {4.0.1},
  doi          = {10.5281/zenodo.17662642},
  url          = {https://doi.org/10.5281/zenodo.17662642},
  swhid        = {swh:1:dir:91ab3394e37a7172edb226c77d86f28d0908c145
                   ;origin=https://doi.org/10.5281/zenodo.8341953;vis
                   it=swh:1:snp:95ebb25732f9dff312b8bb92f923ce4d24d85
                   73f;anchor=swh:1:rel:a0b03c0538a022abf2044a29ca337
                   cf5a61ea308;path=AnMoreau-PyMoosh-be8ac79
                  },
}

@article{baranov2017coherent,
  title={Coherent perfect absorbers: linear control of light with light},
  author={Baranov, Denis G and Krasnok, Alex and Shegai, Timur and Al{\`u}, Andrea and Chong, Yidong},
  journal={Nature Reviews Materials},
  volume={2},
  number={12},
  pages={1--14},
  year={2017},
  publisher={Nature Publishing Group}
}

@article{bennet2024illustrated,
  title={Illustrated tutorial on global optimization in nanophotonics},
  author={Bennet, Pauline and Langevin, Denis and Essoual, Chaymae and Khaireh-Walieh, Abdourahman and Teytaud, Olivier and Wiecha, Peter and Moreau, Antoine},
  journal={JOSA B},
  volume={41},
  number={2},
  pages={A126--A145},
  year={2024},
  publisher={Optica Publishing Group}
}

@article{bhatia2012helmholtz,
  title={The Helmholtz-Hodge decomposition—a survey},
  author={Bhatia, Harsh and Norgard, Gregory and Pascucci, Valerio and Bremer, Peer-Timo},
  journal={IEEE Transactions on visualization and computer graphics},
  volume={19},
  number={8},
  pages={1386--1404},
  year={2012},
  publisher={IEEE}
}

@article{lemaitre2017interferometric,
  title={Interferometric control of the absorption in optical patch antennas},
  author={Lema{\^\i}tre, Caroline and Centeno, Emmanuel and Moreau, Antoine},
  journal={Scientific Reports},
  volume={7},
  number={1},
  pages={2941},
  year={2017},
  publisher={Nature Publishing Group UK London}
}

@article{benedicto2015numerical,
  title={Numerical tool to take nonlocal effects into account in metallo-dielectric multilayers},
  author={Benedicto, Jessica and Poll{\`e}s, R{\'e}mi and Cirac{\`\i}, Cristian and Centeno, Emmanuel and Smith, David R and Moreau, Antoine},
  journal={JOSA A},
  volume={32},
  number={8},
  pages={1581--1588},
  year={2015},
  publisher={Optica Publishing Group}
}

@article{barry2020evolutionary,
  title={Evolutionary algorithms converge towards evolved biological photonic structures},
  author={Barry, Mamadou Aliou and Berthier, Vincent and Wilts, Bodo D and Cambourieux, Marie-Claire and Bennet, Pauline and Poll{\`e}s, R{\'e}mi and Teytaud, Olivier and Centeno, Emmanuel and Biais, Nicolas and Moreau, Antoine},
  journal={Scientific reports},
  volume={10},
  number={1},
  pages={12024},
  year={2020},
  publisher={Nature Publishing Group UK London}
}

@article{mortensen2014generalized,
  title={A generalized non-local optical response theory for plasmonic nanostructures},
  author={Mortensen, N Asger and Raza, S{\o}ren and Wubs, Martijn and S{\o}ndergaard, Thomas and Bozhevolnyi, Sergey I},
  journal={Nature communications},
  volume={5},
  number={1},
  pages={3809},
  year={2014},
  publisher={Nature Publishing Group UK London}
}

@article{vasanelli2020semiconductor,
  title={Semiconductor quantum plasmonics},
  author={Vasanelli, Angela and Huppert, Simon and Haky, Andrew and Laurent, Thibault and Todorov, Yanko and Sirtori, Carlo},
  journal={Physical Review Letters},
  volume={125},
  number={18},
  pages={187401},
  year={2020},
  publisher={APS}
}

@article{halevi1995hydrodynamic,
  title={Hydrodynamic model for the degenerate free-electron gas: generalization to arbitrary frequencies},
  author={Halevi, P},
  journal={Physical Review B},
  volume={51},
  number={12},
  pages={7497},
  year={1995},
  publisher={APS}
}

@article{tonks1929oscillations,
  title={Oscillations in ionized gases},
  author={Tonks, Lewi and Langmuir, Irving},
  journal={Physical Review},
  volume={33},
  number={2},
  pages={195},
  year={1929},
  publisher={APS}
}

@article{melnyk1968resonant,
  title={Resonant excitation of plasmons in thin films by elecromagnetic waves},
  author={Melnyk, Andrew R and Harrison, Michael J},
  journal={Physical Review Letters},
  volume={21},
  number={2},
  pages={85},
  year={1968},
  publisher={APS}
}

@article{anderegg1971optically,
  title={Optically excited longitudinal plasmons in potassium},
  author={Anderegg, M and Feuerbacher, B and Fitton, B},
  journal={Physical Review Letters},
  volume={27},
  number={23},
  pages={1565},
  year={1971},
  publisher={APS}
}

@article{melnyk1970theory,
  title={Theory of optical excitation of plasmons in metals},
  author={Melnyk, Andrew R and Harrison, Michael J},
  journal={Physical Review B},
  volume={2},
  number={4},
  pages={835},
  year={1970},
  publisher={APS}
}

@article{taliercio2014brewster,
  title={Brewster “mode” in highly doped semiconductor layers: an all-optical technique to monitor doping concentration},
  author={Taliercio, Thierry and Guilengui, Vilianne Ntsame and Cerutti, Laurent and Tourni{\'e}, Eric and Greffet, Jean-Jacques},
  journal={Optics express},
  volume={22},
  number={20},
  pages={24294--24303},
  year={2014},
  publisher={Optica Publishing Group}
}

@article{ferrell1962plasma,
  title={Plasma resonance in the electrodynamics of metal films},
  author={Ferrell, Richard A and Stern, Edward A},
  journal={Journal of Quantitative Spectroscopy and Radiative Transfer},
  volume={2},
  number={4},
  pages={679--682},
  year={1962},
  publisher={Elsevier}
}

@article{schmitt2016dgtd,
  title={A DGTD method for the numerical modeling of the interaction of light with nanometer scale metallic structures taking into account non-local dispersion effects},
  author={Schmitt, Nikolai and Scheid, Claire and Lanteri, St{\'e}phane and Moreau, Antoine and Viquerat, Jonathan},
  journal={Journal of Computational Physics},
  volume={316},
  pages={396--415},
  year={2016},
  publisher={Elsevier}
}

@article{kwiecien2023nonlocal,
  title={Nonlocal Fourier modal method for analyzing nonlocal plasmonic periodic nanostructures},
  author={Kwiecien, Pavel and Burda, Milan and Richter, Ivan},
  journal={JOSA B},
  volume={40},
  number={3},
  pages={491--500},
  year={2023},
  publisher={Optica Publishing Group}
}

@article{ajib2019energy,
  title={The energy point of view in plasmonics},
  author={Ajib, Rabih and Pitelet, Armel and Poll{\`e}s, R{\'e}mi and Centeno, Emmanuel and Ajaltouni, Ziad and Moreau, Antoine},
  journal={JOSA B},
  volume={36},
  number={4},
  pages={1150--1154},
  year={2019},
  publisher={Optical Society of America}
}

@article{de2018viscoelastic,
  title={Viscoelastic optical nonlocality of low-loss epsilon-near-zero nanofilms},
  author={De Ceglia, Domenico and Scalora, Michael and Vincenti, Maria A and Campione, Salvatore and Kelley, Kyle and Runnerstrom, Evan L and Maria, Jon-Paul and Keeler, Gordon A and Luk, Ting S},
  journal={Scientific reports},
  volume={8},
  number={1},
  pages={1--11},
  year={2018},
  publisher={Nature Publishing Group}
}

@Article{Ciraci2012,

  author={Cirac{\`\i}, Cristian and Hill, RT and Mock, JJ and Urzhumov, Yaroslav and Fern{\'a}ndez-Dom{\'\i}nguez, AI and Maier, SA and Pendry, JB and Chilkoti, Ashutosh and Smith, DR},
  title   = {Probing the Ultimate Limits of Plasmonic Enhancement},
  journal = {Science},
  year    = {2012},
  volume  = {337},
  number  = {6098},
  pages   = {1072-1074}
}

@Article{Raza2013,
  author    = {Søren Raza and Nicolas Stenger and Shima Kadkhodazadeh and Søren V. Fischer and Natalie Kostesha and Antti-Pekka Jauho and Andrew Burrows and Martijn Wubs and N. Asger Mortensen},
  title     = {Blueshift of the surface plasmon resonance in silver nanoparticles studied with EELS},
  journal   = {Nanophotonics},
  year      = {2013},
  volume    = {2},
  number    = {2},
  pages     = {131 - 138},
  address   = {Berlin, Boston},
  doi       = {https://doi.org/10.1515/nanoph-2012-0032},
  publisher = {De Gruyter},
  url       = {https://www.degruyter.com/view/journals/nanoph/2/2/article-p131.xml},
}

@Article{Raza2015,
  author    = {S{\o}ren Raza and Sergey I Bozhevolnyi and Martijn Wubs and N Asger Mortensen},
  title     = {Nonlocal optical response in metallic nanostructures},
  journal   = {Journal of Physics: Condensed Matter},
  year      = {2015},
  volume    = {27},
  number    = {18},
  pages     = {183204},
  month     = {apr},
  doi       = {10.1088/0953-8984/27/18/183204},
  publisher = {{IOP} Publishing},
  url       = {https://doi.org/10.1088%2F0953-8984%2F27%2F18%2F183204},
}

@Article{Toscano2012,
  author    = {Giuseppe Toscano and S{\o}ren Raza and Antti-Pekka Jauho and N. Asger Mortensen and Martijn Wubs},
  title     = {Modified field enhancement and extinction by plasmonic nanowire dimers due to nonlocal response},
  journal   = {Opt. Express},
  year      = {2012},
  volume    = {20},
  number    = {4},
  pages     = {4176--4188},
  month     = {Feb},
  doi       = {10.1364/OE.20.004176},
  publisher = {OSA},
  url       = {http://www.opticsexpress.org/abstract.cfm?URI=oe-20-4-4176},
}

@Article{Ciraci2013,
  author   = {Ciracì, Cristian and Pendry, John B. and Smith, David R.},
  title    = {Hydrodynamic Model for Plasmonics: A Macroscopic Approach to a Microscopic Problem},
  journal  = {ChemPhysChem},
  year     = {2013},
  volume   = {14},
  number   = {6},
  pages    = {1109-1116},
  doi      = {10.1002/cphc.201200992},
  url      = {https://chemistry-europe.onlinelibrary.wiley.com/doi/abs/10.1002/cphc.201200992},
}

@Article{Moreau2013,
  author    = {Moreau, A. and Cirac\`{\i}, C. and Smith, D. R.},
  title     = {Impact of nonlocal response on metallodielectric multilayers and optical patch antennas},
  journal   = {Phys. Rev. B},
  year      = {2013},
  volume    = {87},
  pages     = {045401},
  month     = {Jan},
  doi       = {10.1103/PhysRevB.87.045401},
  issue     = {4},
  numpages  = {11},
  publisher = {American Physical Society},
  url       = {https://link.aps.org/doi/10.1103/PhysRevB.87.045401},
}

@article{cramer2012numerical,
  title={Numerical estimates for the bulk viscosity of ideal gases},
  author={Cramer, Mark S},
  journal={Physics of fluids},
  volume={24},
  number={6},
  year={2012},
  publisher={AIP Publishing}
}

@article{burda2023nonlocal,
  title={Nonlocal response of planar plasmonic layers},
  author={Burda, Milan and Richter, Ivan and Kwiecien, Pavel},
  journal={Optical and Quantum Electronics},
  volume={55},
  number={14},
  pages={1286},
  year={2023},
  publisher={Springer}
}

@article{raza2011unusual,
  title={Unusual resonances in nanoplasmonic structures due to nonlocal response},
  author={Raza, S{\o}ren and Toscano, Giuseppe and Jauho, Antti-Pekka and Wubs, Martijn and Mortensen, N Asger},
  journal={Physical Review B—Condensed Matter and Materials Physics},
  volume={84},
  number={12},
  pages={121412},
  year={2011},
  publisher={APS}
}

@article{casias2019carrier,
  title={Carrier concentration and transport in Be-doped InAsSb for infrared sensing applications},
  author={Casias, Lilian K and Morath, Christian P and Steenbergen, Elizabeth H and Webster, Preston T and Kim, Jin K and Cowan, Vincent M and Balakrishnan, Ganesh and Krishna, Sanjay},
  journal={Infrared Physics \& Technology},
  volume={96},
  pages={184--191},
  year={2019},
  publisher={Elsevier}
}

@article{rogalski2020inassb,
  title={InAsSb-based infrared photodetectors: Thirty years later on},
  author={Rogalski, Antoni and Martyniuk, Piotr and Kopytko, Malgorzata and Madejczyk, Pawel and Krishna, Sanjay},
  journal={Sensors},
  volume={20},
  number={24},
  pages={7047},
  year={2020},
  publisher={MDPI}
}

@article{wieder1974transport,
  title={Transport coefficients of InAs epilayers},
  author={Wieder, HH},
  journal={Applied Physics Letters},
  volume={25},
  number={4},
  pages={206},
  year={1974}
}

@article{conti1999elasticity,
  title={Elasticity of an electron liquid},
  author={Conti, Sergio and Vignale, Giovanni},
  journal={Physical Review B},
  volume={60},
  number={11},
  pages={7966},
  year={1999},
  publisher={APS}
}

@article{ciraci2017current,
  title={Current-dependent potential for nonlocal absorption in quantum hydrodynamic theory},
  author={Cirac{\`\i}, Cristian},
  journal={Physical Review B},
  volume={95},
  number={24},
  pages={245434},
  year={2017},
  publisher={APS}
}
\end{document}